\documentclass[reqno]{amsart}                     
\usepackage{graphicx}
\usepackage[T1]{fontenc}
\usepackage[utf8]{inputenc}
\usepackage{geometry}
\usepackage{pifont}
\usepackage{amsmath}
\usepackage{amssymb}
\usepackage[all]{xy}
\newcommand{\ray}[1]{\mathsf{#1}}
\newcommand{\trafo}[1]{\mathbf{#1}}

\newcommand{\pr}{\odot}

\begin{document}

\title{On the realization of Symmetries in Quantum Mechanics}
\author{Kai Johannes Keller, Nikolaos A. Papadopoulos and  Andr\'es F. Reyes-Lega}
\thanks{\hspace{-0.45cm}Kai Johannes Keller\\
Inst. f. Physik (WA THEP) der Johannes Gutenberg-Universit{\"a}t Mainz, Germany\\
E-mail: kai.johannes.keller@desy.de\\
\emph{Present address:} II. Inst. f. Theoretische Physik der Universit{\"a}t Hamburg
\vspace{0.15cm}\\
\hspace{-0.45cm}Nikolaos A. Papadopoulos\\
Inst. f. Physik (WA THEP) der Johannes Gutenberg-Universit{\"a}t Mainz, Germany\\
E-mail: papadopoulos@thep.physik.uni-mainz.de\vspace{0.15cm}\\
\hspace{-0.45cm}Andr\'es F. Reyes-Lega\\
Departamento de F\'isica, Universidad de los Andes, Bogot\'a,
Colombia\\
E-mail: anreyes@uniandes.edu.co}

\begin{abstract}
The aim of this paper is to give a simple, geometric proof of Wigner's
theorem on the realization of symmetries in quantum mechanics that
clarifies its relation to projective geometry. Although several proofs exist already,
it seems that the relevance of Wigner's theorem is
not fully appreciated in general. It is Wigner's theorem which allows
the use of linear realizations of symmetries and therefore guarantees
that, in the end, quantum theory stays a linear theory. In the present
paper, we take a strictly geometrical point of view in order to prove
this theorem. It becomes apparent that Wigner's theorem is nothing
else but a corollary of the fundamental theorem of projective geometry.
In this sense, the proof presented here is simple, transparent and
therefore accessible even to elementary treatments in quantum mechanics.
\end{abstract}
\maketitle
\section{\label{sec:Introduction}Introduction}

There is no doubt that symmetries play a very important role in physics.
For any given problem the existence of symmetries is decisive and
their realization is crucial when it comes to the solution of the
problem. How symmetries are realized depends of course on the theory
under consideration and more precisely on the corresponding structure
of its space of states. It is well known that in quantum mechanics
symmetries are realized by special linear or anti-linear operators.
We expect this since the Hilbert space is linear. In this consideration,
however, we overlook the fact that, because of the probability interpretation
of the wave function, a state is not a single vector but a ray in Hilbert
space, i.e. a one-dimensional subspace. Consequently, the space of
quantum mechanical states is not a linear but a projective space.
Since projective spaces are not linear (and hence the space of states
is not linear) the linear realization of symmetries cannot be taken
for granted. On the contrary we may doubt if such a linear realization
is possible at all. It is due to the work of E. Wigner and his almost
forgotten theorem on the realization of symmetries, a theorem disregarded even
by many otherwise excellent and generally accepted books and lectures on
quantum mechanics, that we may nevertheless always use linear or anti-linear
realizations in quantum mechanics. It states that\vspace{0.2cm}
\begin{center}
\begin{tabular}{|p{14.6cm}|}
\hline
\emph{Any symmetry transformation on the set of pure states of a quantum
mechanical system is represented up to a scalar factor by either a
unitary or an anti-unitary transformation on the corresponding Hilbert
space.}
\\
 \hline
\end{tabular}
\\
\end{center}
\vspace{0.2cm}
By symmetry transformation we mean here a bijective map,
a ray transformation which preserves transition probabilities.

We intend to discuss this so often ignored theorem and to give an
elementary and simple proof, which relies on its geometrical background.
This background of course is dominated by projective geometry. Despite
its non linear nature it is projective geometry that gives a deeper
justification for the usage of linear or anti-linear realizations
in quantum mechanics and consequently in quantum field theory. In
the literature one can find a variety of different proofs of Wigner's
theorem: \cite{Wigner1931,Hagedorn1959,Uhlhorn1962,Emch1963,Lomont1963,Bargmann1964,Varadarajan1968,Bracci1975,Weinberg1995,Cassinelli1997}
and there are some that make a connection to projective geometry: \cite{Uhlhorn1962,Emch1963,Lomont1963,Varadarajan1968}. However, almost all of
them make no direct use of it, and where they do the key concept is
covered by a complex presentation or by a generalizing idea in one
or the other direction. The aim of this paper is to give a simple
geometric proof of Wigner's original theorem without any generalization
whatsoever, and as a consequence to clarify the key role projective
geometry has when it comes to the representation of symmetries within
quantum physics.

In this paper the proof of Wigner's theorem is a direct consequence
of projective geometry: Wigner's theorem is essentially a corollary
of the fundamental theorem of projective geometry. In this sense, Wigner's
theorem could be treated in an elementary way in any quantum mechanics
textbook and subsequently in any lecture.

In the following we will first give a short review of the geometric proof
of Wigner's theorem. The proof contains four steps. We first show
that a symmetry transformation fulfills the premises of the fundamental
theorem of projective geometry: any symmetry transformation is a collineation,
i.e. it maps projective lines to projective lines. In the second step
we use the fundamental theorem of projective geometry in the form
and proof given by E. Artin \cite{Artin1957}: any collineation between
finite dimensional projective spaces over an arbitrary field $\mathbb{K}$
is a semi-projectivity. In this paper we use only $\mathbb{K}\in\left\{ \mathbb{R},\mathbb{C}\right\} $,
denoting either the field of real or the field of complex numbers.
By semi-projectivity we mean a bijective map between projective spaces
which is induced by a bijective linear or anti-linear map between
the associated vector spaces. Observe that the fundamental theorem
of projective geometry refers to (finite-dimensional) $\mathbb{K}$-vector
spaces with no further structure and not to Hilbert spaces so that
we cannot talk about unitarity here. In the third step we take a Hilbert
space instead of a vector space and we consider the corresponding
formulation as established in Artin's proof. In this way we can show that the
linear operator is a unitary and the anti-linear operator is an anti-unitary
map. In the last step the extension to infinite dimensions is performed.
One can imagine that the difficulty is to prove the existence of a
ray consistent semi-linear transformation, which is done in Artin's
proof in the second step. What the other steps are concerned with
is quite easy to prove as soon as you have realized the existence
of the corresponding statements.

The remainder of this paper is organized as follows: in the subsequent
section (\ref{sec:fundamental-theorem-projective-geometry}) we discuss
the basic concepts of projective geometry and introduce its fundamental
theorem. We refer very shortly to the first proof given by F. Klein \cite{Klein1925},
who also used concepts found by A. F. Möbius, but for our treatment
we rely mainly on the work of E. Artin. We summarize briefly the steps
of his proof since we find it useful for the understanding of the
topic. We will argue how this is related to Wigner's theorem in section
\ref{sec:Symmetries-and-Wigners-theorem}. In section \ref{sec:Geometric-proof}
we give our proof of Wigner's theorem, which is, within the framework
of projective geometry, simple and elementary. The key points are
summarized and a short conclusion is given in the last section (\ref{sec:Conclusion}).

\section{\label{sec:fundamental-theorem-projective-geometry}On the fundamental
theorem of projective geometry}

Before we start with the discussion of Wigner's theorem and its geometric
proof, we use this preliminary section to clarify the notation used
throughout the paper.

A \emph{ray} $\ray{A}$ in projective geometry is an orbit $\left[a\right]$
of the group of units $\mathbb{K}^{\times}$ on a vector space $V$
over the field $\mathbb{K}$, i.e. $\ray{A}$ is a one-dimensional
subspace of $V$ (with the zero element removed). We write $\left[a\right]$
for the ray with representative $a\in V$. The set of all such rays
in $V$ is called \emph{projective space} $\mathcal{P}V$. Its dimension
as a manifold is
\begin{equation}
\dim\mathcal{P}V=\dim V-1\,.\label{eq:dim}
\end{equation}

Some of the properties of vectors in $V$ are preserved as we go to
the projection $\mathcal{P}V$. We call a set $\left\{ \ray{A}_{1},\dots,\ray{A}_{n}\right\} $
of two or more rays \emph{projectively independent}, if and only if
there is a linearly independent set of vectors $\left\{ a_{1},\dots,a_{n}\right\} $
such that $\forall k\in\left\{ 1,\dots,n\right\} :$ $a_{k}\in\ray{A}_{k}$.
Observe that two rays are either projectively independent or equal.
In a projective space $\mathcal{P}V$ of dimension $n$ a set $\mathcal{B}\equiv\left\{ \ray{B}_{1},\dots,\ray{B}_{n+2}\right\} \subset\mathcal{P}V$
of $n+2$ rays is called a projective base of $\mathcal{P}V$, if
and only if any subset of $\mathcal{B}$ containing $n+1$ rays is
projectively independent.

Regarding two different rays $\ray{A},\ray{B}\in\mathcal{P}V$, there
is a natural operation, the \emph{unification}, defined by:
\begin{equation}
\ray{A}\vee\ray{B}:=\left\{ \left[a+b\right]:\, a\in\ray{A},\, b\in\ray{B}\right\} \,.\label{eq:proj-line}
\end{equation}
$\ray{A}\vee\ray{B}$ is the plane spanned by the projectively independent
rays $\ray{A}$ and $\ray{B}$. Since $\dim\left(\ray{A}\vee\ray{B}\right)=1$
it is natural to refer to this unification as the \emph{projective
line} uniquely determined by the projective points $\ray{A}$ and
$\ray{B}$. This gives rise to another notion used in projective geometry:
three or more distinct points $\ray{A}_{1},\dots,\ray{A}_{n}$ are
called \emph{collinear} if and only if they are on the same projective
line, i.e. $\forall k\in\left\{ 1,\dots,n\right\} :$ $\ray{A}_{k}\in\ray{A}_{1}\vee\ray{A}_{2}$.

With the notions we introduced above, we are now able to define the
basic maps between projective spaces which are the collineation and
the semi-projectivity. A \emph{collineation} $\trafo{f}:\mathcal{P}V\rightarrow\mathcal{P}W$
is a bijective map that preserves collinearity, i.e. $\trafo{f}$
maps any projective line to a projective line:
\begin{equation}
\trafo{f}\left(\ray{A}\vee\ray{B}\right)=\trafo{f}\left(\ray{A}\right)\vee\trafo{f}\left(\ray{B}\right)\,.\label{eq:collineation}
\end{equation}
A \emph{semi-projectivity} $\trafo{g}:\mathcal{P}V\rightarrow\mathcal{P}W$
on the other hand is a bijective map that is induced by a semi-linear
map $G:V\rightarrow W$, i.e.
\begin{equation}
\left[Ga\right]=\trafo{g}\left(\left[a\right]\right)\,.\label{eq:semi-projectivity-compatible}
\end{equation}
Any map $G$ that fulfills equation (\ref{eq:semi-projectivity-compatible})
is called \emph{compatible} with $\trafo{g}$. The prefix {}``semi''
stands for {}``up to a field automorphism''. Hence the semi-linear
map $G$ is linear up to a field automorphism $\sigma:\mathbb{K}\rightarrow\mathbb{K}$,
i.e. $\forall\alpha,\beta\in\mathbb{K}$, $\forall a,b\in V$:
\begin{equation}
G\left(\alpha a+\beta b\right)=\sigma\left(\alpha\right)\, Ga+\sigma\left(\beta\right)\, Gb\,.\label{eq:semi-linear}
\end{equation}
Later it will become apparent that the only field automorphisms occuring
in the context of Wigner's theorem are either the identity or complex
conjugation. Hence, the map $G$ is either linear or anti-linear.
Obviously, any semi-projectivity preserves collinearity, i.e. is a
collineation. The reverse statement is also true, but not trivial
at all, it is the \emph{fundamental theorem of projective geometry}:
\vspace{0.2cm}
\begin{center}
\begin{tabular}{|p{14.6cm}|}
\hline
\emph{\label{fundamental-theorem-of-projective-geometry}Let $\mathcal{P}V$
and $\mathcal{P}W$ be projective spaces of same dimension $n\geq2$.
Then any collineation $\trafo{f}:\mathcal{P}V\rightarrow\mathcal{P}W$
is a semi-projectivity, i.e. the following diagram commutes:}

\begin{center}
$\xymatrix{V\ar@{->>}[d]\ar@{-->}[r]^{F} & W\ar@{->>}[d]\\
\mathcal{P}V\ar@{->}[r]^{\trafo{f}} & \mathcal{P}W}
$
\par\end{center}

where $\trafo{f}$ is a collineation and $F$ is semi-linear.
\\
 \hline
\end{tabular}
\end{center}
\vspace{0.2cm}
There are essentially two ways to prove this theorem, the first of
which was carried out by F. Klein in 1925 \cite{Klein1925}. Klein explicitly
shows the assertion for the real projective plane but his construction
can be generalized to real projective spaces of arbitrary (finite)
dimension (for a more detailed discussion consult for example the
corresponding section (I.2) of reference \cite{Keller2006}).
Out of a projective base and by using only the properties of the collineation
he constructs a dense set of intersection points of projective lines.
Hence there is only one continuous collineation that maps the projective
base to the corresponding image base. But the base and its image also
determine a projectivity, and Klein concludes that the regarded collineation
is a projectivity by the fact that any projectivity also is a collineation.

The second proof was introduced by E. Artin in 1957 \cite{Artin1957}
and is valid for projective spaces of (finite) dimension greater or
equal two over an arbitrary field. The intention is to show that for
every collineation $\trafo{f}:\mathcal{P}V\rightarrow\mathcal{P}W$
there exists a semi-linear transformation $F:V\rightarrow W$ which
is compatible with $\trafo{f}$. This means that for a given basis
$\left\{ v_{k}\right\} _{k\in\left\{ 0,\dots,n\right\} }$ of $V$
there exists a basis $\left\{ w_{k}\right\} _{k\in\left\{ 0,\dots,n\right\} }$
of $W$ and a semi-linear transformation $F$ such that $\forall k\in\left\{ 0,\dots,n\right\} $:
$w_{k}=F\left(v_{k}\right)$. Due to semi-linearity, a given field
automorphism $\sigma:\mathbb{K}\rightarrow\mathbb{K}$, $F$ then
obviously fulfills $\forall\lambda_{k}\in\mathbb{K}$: $F\left(\lambda_{0}v_{0}+\cdots+\lambda_{n}v_{n}\right)=\sigma\left(\lambda_{0}\right)\, w_{0}+\cdots+\sigma\left(\lambda_{n}\right)\, w_{n}$.
Furthermore compatibility requires:
\begin{equation}
\trafo{f}\left[\lambda_{0}v_{0}+\cdots+\lambda_{n}v_{n}\right]=\left[\sigma\left(\lambda_{0}\right)\, w_{0}+\cdots+\sigma\left(\lambda_{n}\right)\, w_{n}\right]\,.\label{eq:Artin-to-show}
\end{equation}
In other words, the image of a representative of a ray in $V$ given
by a linear combination of the basis vectors and expressed with the
{}``same'' (up to a field automorphism) linear combination of the
corresponding basis vectors in $W$, is a representative of the image
ray, given above. Artin's proof of this assertion is divided into
several steps. As one expects, collinearity is used in almost all
of them. Additionally, he uses induction with respect to the dimensional
parameter $n$ in order to generalize his construction to spaces of
arbitrary (finite) dimension. In what follows we review shortly some
of the steps since we believe that this contributes essentially to
the understanding of the topic.

\begin{enumerate}
\item Using induction with respect to the dimension allows to show that
$\trafo{f}$ conserves not only projective lines, but also projective
subspaces of arbitrary, finite dimension, $\forall n\in\mathbb{N}$:\begin{equation}
\trafo{f}\left(\ray{A}_{0}\vee\ray{A}_{1}\vee\cdots\vee\ray{A}_{n}\right)=\trafo{f}\left(\ray{A}_{0}\right)\vee\trafo{f}\left(\ray{A}_{1}\right)\vee\cdots\vee\trafo{f}\left(\ray{A}_{n}\right)\,.\label{eq:Artin-1}\end{equation}

\item For every basis $\left\{ v_{k}\right\} _{k\in\left\{ 0,\dots,n\right\} }$
in $V$ there exists a basis $\left\{ w_{k}\right\} _{k\in\left\{ 0,\dots,n\right\} }$
in $W$ such that $\forall k\in\left\{ 0,\dots,n\right\} $:\begin{equation}
\trafo{f}\left[v_{k}\right]=\left[w_{k}\right]\qquad\text{and}\qquad\trafo{f}\left[v_{0}+v_{k}\right]=\left[w_{0}+w_{k}\right]\,.\label{eq:Artin-2}\end{equation}

\item There exists a $\mathbb{K}$-automorphism $\sigma:\mathbb{K}\rightarrow\mathbb{K}$,
such that $\forall k\in\left\{ 1,\dots,n\right\} $, $\lambda\in\mathbb{K}$:\begin{equation}
\trafo{f}\left[v_{0}+\lambda v_{k}\right]=\left[w_{0}+\sigma\left(\lambda\right)w_{k}\right]\,.\label{eq:Artin-3}\end{equation}

\item Using again induction, one can show that:\begin{equation}
\trafo{f}\left[v_{0}+\lambda_{1}v_{1}+\cdots+\lambda_{n}v_{n}\right]=\left[w_{0}+\sigma\left(\lambda_{1}\right)\, w_{1}+\cdots+\sigma\left(\lambda_{n}\right)\, w_{n}\right]\,.\label{eq:Artin-4}\end{equation}

\item This statement, by using the fact that $\sigma$ is a field automorphism,
finally leads to\begin{equation}
\trafo{f}\left[\lambda_{0}v_{0}+\lambda_{1}v_{1}+\cdots+\lambda_{n}v_{n}\right]=\left[\sigma\left(\lambda_{0}\right)\, w_{0}+\sigma\left(\lambda_{1}\right)\, w_{1}+\cdots+\sigma\left(\lambda_{n}\right)\, w_{n}\right]\,,\label{eq:Artin-5}\end{equation}
which completes Artin's proof of the fundamental theorem of projective
geometry.
\end{enumerate}

Let us just make one small remark on the way the field automorphism
$\sigma:\mathbb{K}\rightarrow\mathbb{K}$ that we have not found in
the proof by F. Klein enters in the third step of Artin's proof.
To be a bit more precise, Artin introduces maps
\[
\begin{array}{ccl}
\lambda & \mapsto & \left[v_{0}+\lambda v_{k}\right]\\
\mathbb{K} & \rightarrow & \left[v_{0}\right]\vee\left[v_{k}\right]
\end{array}
\]
which are just one way to write the homeomorphisms between the projective
lines $\left[v_{0}\right]\vee\left[v_{k}\right]$ and the one-point
compactification $\mathbb{K}\cup\left\{ \infty\right\} $ of the field
$\mathbb{K}$. He gets seemingly different maps $\sigma_{k}:\mathbb{K}\rightarrow\mathbb{K}$
by applying the following diagram

\begin{center}
$\xymatrix{\mathbb{K}\cup\{\infty\}\ar@{^{(}->>}[d]\ar@{-->}[r]^{\sigma_{i}} & \mathbb{K}\cup\{\infty\}\\
\left[v_{0}\right]\vee\left[v_{i}\right]\ar@{^{(}->>}[r]^{\trafo{f}} & \left[w_{0}\right]\vee\left[w_{i}\right]\ar@{^{(}->>}[u]}
$
\par\end{center}

But the maps $\sigma_{k}$ can then be shown to be all the same, unique
field automorphism $\sigma:\mathbb{K}\rightarrow\mathbb{K}$.

\section{\label{sec:Symmetries-and-Wigners-theorem}Symmetries and Wigner's
theorem}

It is well known that the set of pure states of a quantum mechanical
system may be described by the set of one-dimensional subspaces of
the corresponding Hilbert space, i.e. the projective Hilbert space.

If we define the projective space for some Hilbert space $\left(\mathcal{H},\left\langle \cdot|\cdot\right\rangle \right)$,
we find an additional structure on $\mathcal{PH}$, stemming from
the scalar product $\left\langle \cdot|\cdot\right\rangle :\mathcal{H}\times\mathcal{H}\rightarrow\mathbb{C}$
on $\mathcal{H}$. A natural definition for this structure is:\begin{equation}
\ray{A}\pr\ray{B}:=\frac{\left|\left\langle a|b\right\rangle \right|^{2}}{\left\Vert a\right\Vert ^{2}\left\Vert b\right\Vert ^{2}}\quad\forall a\in\ray{A},\, b\in\ray{B}\,.\label{eq:transition-probability}\end{equation}
This is the transition probability between states $\ray{A}$ and $\ray{B}$.
Observe that if $\ray{A}\pr\ray{B}=0$ any representative of $\ray{A}$
is orthogonal to any representative of $\ray{B}$. On the other hand,
if $\ray{A}\pr\ray{B}=1$ then, by Schwarz's inequality $\ray{A}=\ray{B}$.

We define the \emph{projective Hilbert space} to be the pair $\left(\mathcal{PH},\pr\right)$,
where $\mathcal{PH}$ is the ordinary projective space corresponding
to the vector space structure and $\pr:\mathcal{PH}\times\mathcal{PH}\rightarrow\left[0,1\right]$
is the function induced by the scalar product, defined as above.

A symmetry transformation is a map on the space of states that preserves
transition probabilities. Hence, within our notation a symmetry transformation
$\trafo{T}$ is a bijective map $\trafo{T}:\mathcal{PH}\rightarrow\mathcal{PH}'$
such that $\forall\ray{A},\ray{B}\in\mathcal{PH}$: \begin{equation}
\ray{A}\pr\ray{B}=\trafo{T}\ray{A}\pr\trafo{T}\ray{B}\,.\label{eq:quasi-unitarity}\end{equation}
Obviously, by this equation $\trafo{T}$ preserves orthogonality,
and this is why we may also call this property of the symmetry transformation
$\trafo{T}$ \emph{quasi-unitarity}.

With the help of these refined notions we can restate Wigner's theorem
in a more mathematical way.

\emph{\label{fundamental-theorem-of-projective-geometry}Let $\mathcal{P}V$
and $\mathcal{P}W$ be projective spaces of same dimension $n\geq2$.
Then any collineation $\trafo{f}:\mathcal{P}V\rightarrow\mathcal{P}W$
is a semi-projectivity, i.e. the following diagram commutes:}

\begin{center}
$\xymatrix{V\ar@{->>}[d]\ar@{-->}[r]^{F} & W\ar@{->>}[d]\\
\mathcal{P}V\ar@{->}[r]^{\trafo{f}} & \mathcal{P}W}
$
\par\end{center}
\noindent
where $\trafo{f}$ is a collineation and $F$ is semi-linear.
\begin{center}
\begin{tabular}{|p{14.6cm}|}
\hline
\\
\emph{Let $\trafo{T}:\mathcal{PH}\rightarrow\mathcal{PH}'$ be a symmetry
(i.e. quasi-unitary) transformation, then there exists a compatible
semi-unitary transformation $U:\mathcal{H}\rightarrow\mathcal{H}'$,
i.e. $\forall a\in\mathcal{H}$:
\begin{equation*}
\trafo{T}\left[a\right]=\left[Ua\right]
\end{equation*}
and hence the following diagram commutes}:\[
\xymatrix{\mathcal{H}\ar@{->>}[d]\ar@{-->}[r]^{U} & \mathcal{H}'\ar@{->>}[d]\\
\mathcal{PH}\ar@{->}[r]^{\trafo{T}} & \mathcal{PH}'}
\]
\\
 \hline
\end{tabular}
\end{center}
\vspace{0.2cm}

What we encounter here is obviously a lifting problem. Observe the
similarity to the fundamental theorem of projective geometry. It is
exactly this similarity that we will use in order to formulate a geometric
proof of Wigner's theorem. Despite this observation, it is not a priori
clear that this apparent similarity indeed leads to a geometric proof
of Wigner's theorem, which, as we will see in the following section,
it does.

We know that the set of pure states is divided into disjoint super
selection sectors. Due to the super selection rules states of different
super selection sectors cannot be superposed. Nevertheless a symmetry
may be a map from one of these sectors to another. In order to be
able to formulate and prove the theorem without further difficulties,
the regarded projective Hilbert spaces should be two (or even the
same) super selection sectors, in which the superposition principle
holds without limitations.

\section{\label{sec:Geometric-proof}Geometric proof of Wigner's theorem}

The proof we present here consists of four steps. In the first step
we will show that any quantum symmetry transformation is also a projective
collineation. This is the key that makes it possible to apply the
fundamental theorem of projective geometry. In the second step, using
the fundamental theorem of projective geometry, we show that any symmetry
transformation (which is, as shown in step 1 below, a collineation)
is induced by a semi-linear transformation. In step 3 we show
that this semi-linear transformation is actually semi-unitary. This
completes our proof of Wigner's theorem for finite dimensional Hilbert
spaces. In the last step we extend this result to Hilbert spaces of
infinite dimension.

\subsection*{Step 1}
$\,$

\begin{center}
\begin{tabular}{|p{14.6cm}|}
\hline
\begin{center}
\emph{Any symmetry (quasi-unitary) transformation is a collineation.}
\end{center}
\\
 \hline
\end{tabular}
\end{center}

We have to show that \begin{equation}
\trafo{T}\left(\ray{A}\vee\ray{B}\right)=\trafo{T}\ray{A}\vee\trafo{T}\ray{B}\label{eq:symmetry-transformation-collinear-to-show}\end{equation}
for any quasi-unitary transformation $\trafo{T}$. To prove this,
we take a projective line $\ray{A}\vee\ray{B}$ and choose the rays
$\ray{A}$ and $\ray{B}$ to be orthogonal. Then there is an orthogonal
base (OGB) $\left\{ b_{k}\right\} _{k\in I}$ of $\mathcal{H}$, such
that\begin{equation}
\left[b_{1}\right]=\ray{A}\quad\mbox{and}\quad\left[b_{2}\right]=\ray{B}\,.\label{eq:proof-coll-1}\end{equation}
In this base we can write any representative $c$ of any ray $\ray{C}\in\ray{A}\vee\ray{B}$
as\begin{equation}
c=\gamma_{1}b_{1}+\gamma_{2}b_{2}\,.\label{eq:proof-coll-2}\end{equation}
Obviously $\ray{A}\pr\ray{B}=0$, and because $\trafo{T}$ is quasi-unitary
we also have\begin{equation}
\trafo{T}\ray{A}\pr\trafo{T}\ray{B}=0\,.\label{eq:proof-coll-3}\end{equation}
Furthermore $\trafo{T}$ maps the orthogonal rays $\left\{ \left[b_{k}\right]\right\} _{k\in I}$
onto orthogonal rays $\left\{ \left[b_{k}'\right]\right\} _{k\in I}$.
This is elementary since for $\ray{B}_{k}=\left[b_{k}\right]$ and
$\ray{B}_{k}'=\left[b_{k}'\right]$ we have\begin{equation}
\ray{B}_{k}'\pr\ray{B}_{l}'=\trafo{T}\ray{B}_{k}\pr\trafo{T}\ray{B}_{l}=\ray{B}_{k}\pr\ray{B}_{l}=\delta_{kl}\,.\label{eq:ogr-to-ogr}\end{equation}
Since $\trafo{T}$ is bijective, any set of representatives $\left\{ b_{k}'\right\} _{k\in I}$
of these orthogonal rays $\left\{ \ray{B}_{k}'\right\} _{k\in I}$
is an orthogonal base of $\mathcal{H}'$. Hence we may write any representative
$c'$ of $\trafo{T}\ray{C}\in\trafo{T}\left(\ray{A}\vee\ray{B}\right)$
as\begin{equation}
c'=\sum_{k\in I}\gamma_{k}'\: b_{k}'\,.\label{eq:proof-coll-4}\end{equation}
But for the coefficients $\gamma_{k}'$ we have\begin{equation}
\frac{\left\Vert b_{k}'\right\Vert ^{2}}{\left\Vert c'\right\Vert ^{2}}\left|\gamma_{k}'\right|^{2}=\frac{\left|\left\langle b_{k}'|c'\right\rangle \right|^{2}}{\left\Vert b_{k}'\right\Vert ^{2}\left\Vert c'\right\Vert ^{2}}=\trafo{T}\left[b_{k}\right]\pr\trafo{T}\left[c\right]=\left[b_{k}\right]\pr\left[c\right]=\frac{\left|\left\langle b_{k}|c\right\rangle \right|^{2}}{\left\Vert b_{k}\right\Vert ^{2}\left\Vert c\right\Vert ^{2}}=\frac{\left\Vert b_{k}\right\Vert ^{2}}{\left\Vert c\right\Vert ^{2}}\left|\gamma_{k}\right|^{2},\label{eq:proof-coll-5}\end{equation}
and hence $\gamma_{k}'=0$ $\forall k\geq3$. Equation (\ref{eq:proof-coll-4})
then reduces to\begin{equation}
c'=\gamma_{1}'\: b_{1}'+\gamma_{2}'\: b_{2}'\,.\label{eq:proof-coll-6}\end{equation}
This means that any representative of $\trafo{T}\ray{C}\in\trafo{T}\left(\ray{A}\vee\ray{B}\right)$
is an element of the plane spanned by the rays $\left[b_{1}'\right]=\trafo{T}\ray{A}$
and $\left[b_{2}'\right]=\trafo{T}\ray{B}$, i.e. the projective line
$\trafo{T}\ray{A}\vee\trafo{T}\ray{B}$.

Hence it follows that \begin{equation}
\trafo{T}\left(\ray{A}\vee\ray{B}\right)\subset\trafo{T}\ray{A}\vee\trafo{T}\ray{B}\,.\label{eq:proof-coll-7}\end{equation}
And again since $\trafo{T}$ is bijective one easily verifies that:

\begin{equation}
\trafo{T}\left(\ray{A}\vee\ray{B}\right)=\trafo{T}\ray{A}\vee\trafo{T}\ray{B}\,.\label{eq:symmetry-transformation-collinear}\end{equation}
Thus any quasi-unitary transformation $\trafo{T}$ is a collineation.

\subsection*{Step 2}

Since we know that $\trafo{T}$ is a collineation, we can apply the
fundamental theorem of projective geometry (p. \pageref{fundamental-theorem-of-projective-geometry}).
It then follows that $\trafo{T}$ is a semi-projectivity. This means
that $\trafo{T}$ is induced by either a linear or an anti-linear
transformation $U$ between the $\mathbb{C}$-vector spaces $\mathcal{H}$
and $\mathcal{H}'$, which conversely is compatible with the ray transformation
$\trafo{T}$:

\begin{center}
\begin{tabular}{|p{14.6cm}|}
\hline
\begin{center}
\emph{Any symmetry (quasi-unitary) transformation $\trafo{T}$}\\
\emph{is induced by a semi-linear transformation $U$:}
\par\end{center}

\[
\xymatrix{\mathcal{H}\ar@{->>}[d]\ar@{-->}[r]_{\text{\tiny{semi-linear}}}^{U} & \mathcal{H}'\ar@{->>}[d]\\
\mathcal{PH}\ar@{->}[r]^{\trafo{T}} & \mathcal{PH}'}
\]
\\
 \hline
\end{tabular}
\end{center}

Observe that in $\mathbb{C}$ there are only two field automorphisms
mapping the unit element to itself: the identity and the complex conjugation.
Apparently, regarding equations (\ref{eq:Artin-2}) and (\ref{eq:Artin-3}),
the field automorphism $\sigma$ fulfills $\sigma(1)=1$. Hence $U$
is either linear or anti-linear.

\subsection*{Step 3}

Now, as implicitly in the first step, too, we consider $\mathcal{H}$
as a Hilbert space. So in addition we dispose of the standard Hermitian
scalar product. The symmetry transformation $\trafo{T}$ conserves
probabilities by assumption:\[
\trafo{T}\ray{A}\pr\trafo{T}\ray{B}=\ray{A}\pr\ray{B}\,.\]
Then, by compatibility, the semi-linear transformation $U$ also respects
transition probabilities. Additionally it maps orthogonal vectors
onto orthogonal vectors. This is enough to demand for $U$ mapping
one ONB to another ONB. More explicitly, we have:\begin{equation}
\left[b_{1}'+b_{k}'\right]\pr\left[b_{1}'\right]=\left[b_{1}+b_{k}\right]\pr\left[b_{1}\right]\label{eq:proof-ONB-1}\end{equation}
From which by $\frac{\left\Vert b_{1}+b_{k}\right\Vert ^{2}\left\Vert b_{1}\right\Vert ^{2}}{\left|\left\langle b_{1}+b_{k}|b_{1}\right\rangle \right|^{2}}=\frac{\left(\left\Vert b_{1}\right\Vert ^{2}+\left\Vert b_{k}\right\Vert ^{2}\right)\left\Vert b_{1}\right\Vert ^{2}}{\left\Vert b_{1}\right\Vert ^{4}}=1+\frac{\left\Vert b_{k}\right\Vert ^{2}}{\left\Vert b_{1}\right\Vert ^{2}}$
it follows that\begin{equation}
\frac{\left\Vert b_{k}'\right\Vert ^{2}}{\left\Vert b_{1}'\right\Vert ^{2}}=\frac{\left\Vert b_{k}\right\Vert ^{2}}{\left\Vert b_{1}\right\Vert ^{2}}\,.\label{eq:proof-ONB-2}\end{equation}
Hence if we choose $\left\{ b_{k}\right\} _{k\in I}$ to be an ONB
of $\mathcal{H}$, we can choose $\left\Vert b_{1}'\right\Vert =1$
and obtain another ONB $\left\{ b_{k}'\right\} _{k\in I}$ of $\mathcal{H}'$.
Then, since any semi-linear transformation is semi-unitary, if it
maps one ONB to another, we conclude that the map $U$, defined by\begin{equation}
Ub_{k}:=b_{k}'\quad\forall k\in I\label{eq:def-U}\end{equation}
is semi-unitary. This completes our proof of Wigner's theorem for
a finite dimensional Hilbert space.

We already know from Artin's proof that the image base $\left\{ b_{k}'\right\} _{k\in I}$
is determined up to one single overall factor, so, since we already
asked for $\left\Vert b_{1}'\right\Vert $ to be one, there's one
last degree of freedom left in the choice of the semi-unitary map
$U$: the semi-unitary map $U$ is determined up to a phase factor.

We have shown Wigner's theorem in a geometric way that emphasizes
and clarifies the connection to the fundamental theorem of projective
geometry at least for finite index sets $I$. Observe that the fundamental
theorem of projective geometry is a theorem that deals with (finite
dimensional) $\mathbb{K}$-vector spaces. In order to obtain Wigner's
theorem for an infinite dimensional Hilbert space, questions of convergence
have to be examined. Notice however that steps 1 and 3 are valid also
for infinite dimensions.

\subsection*{Step 4}

Now $\mathcal{H}$ and $\mathcal{H}'$ are infinite dimensional spaces.
In this case we assume of course that $\trafo{T}$ is continuous.
Taking the bases $\left\{ b_{k}\right\} _{k\in I}$ and $\left\{ b_{k}'\right\} _{k\in I}$
of step 3 as countable bases for some infinite dimensional Hilbert
spaces $\mathcal{H}$ and $\mathcal{H}'$ respectively, we can define
a semi-unitary operator $U:\mathcal{H}\rightarrow\mathcal{H}'$ by:\begin{equation}
Ub_{k}:=b_{k}'\quad\forall k\in I\,.\label{eq:define-U-infinite}\end{equation}
Regardless of the fact that we already used the letter $U$ to denote
the semi-unitary, compatible map in the case of finite dimensional
Hilbert spaces, we use the same letter $U$ to introduce a semi-unitary
map between Hilbert spaces of infinite dimension. The goal of this
last step is to show that the operator $U$, defined by (\ref{eq:define-U-infinite})
is, even in the case of infinite dimensions, compatible with the quantum
symmetry transformation $\trafo{T}:\mathcal{PH}\rightarrow\mathcal{PH}'$,
as the commutative diagram indicates:

\begin{center}
$\xymatrix{\mathcal{H}\ar@{->>}[d]\ar@{-->}[r]^{U} & \mathcal{H}'\ar@{->>}[d]\\
\mathcal{PH}\ar@{->}[r]^{\trafo{T}} & \mathcal{PH}'}
$
\par\end{center}

\noindent This means, we have to show that $\forall\alpha_{k}\in\mathbb{C}$:\begin{equation}
\trafo{T}\left[\sum_{k=1}^{\infty}\alpha_{k}\, b_{k}\right]=\left[\sum_{k=1}^{\infty}\sigma\left(\alpha_{k}\right)\, Ub_{k}\right]\,.\label{eq:infinite-compatibility}\end{equation}

In steps 1-3 we already have proven Wigner's theorem for finite dimensions,
i.e. $\forall n\in\mathbb{N}$:\begin{equation}
\trafo{T}\left[\sum_{k=1}^{n}\alpha_{k}\, b_{k}\right]=\left[\sum_{k=1}^{n}\sigma\left(\alpha_{k}\right)\, Ub_{k}\right]\,.\label{eq:finite-compatibility}\end{equation}
 Since the projection $\pi:\mathcal{H}\ni x\mapsto\left[x\right]\in\mathcal{PH}$
is by definition continuous (with respect to the quotient topology
on $\mathcal{PH}$) we have for all sequences $\left(x_{n}\right)_{n\in\mathbb{N}}$
in $\mathcal{H}$:\begin{equation}
\left[\lim_{n\rightarrow\infty}x_{n}\right]=\lim_{n\rightarrow\infty}\left[x_{n}\right]\,,\label{eq:continuity-projection}\end{equation}
Hence, since the quantum symmetry $\trafo{T}$ is also a continuous
transformation, we obtain directly:\begin{eqnarray}
\lim_{n\rightarrow\infty}\trafo{T}\left[\sum_{k=1}^{n}\alpha_{k}\, b_{k}\right] & = & \lim_{n\rightarrow\infty}\left[\sum_{k=1}^{n}\sigma\left(\alpha_{k}\right)\, Ub_{k}\right]\nonumber \\
\trafo{T}\left[\lim_{n\rightarrow\infty}\sum_{k=1}^{n}\alpha_{k}\, b_{k}\right] & = & \left[\lim_{n\rightarrow\infty}\sum_{k=1}^{n}\sigma\left(\alpha_{k}\right)\, Ub_{k}\right]\,,\label{eq:infinite-compatibility-result}\end{eqnarray}
which is the assertion (\ref{eq:infinite-compatibility}), now proven
to be correct. This completes the proof of Wigner's theorem also for
infinite dimensions. We see that our derivation is geometric, since
it results with the help of some basic considerations directly from
the fundamental theorem of projective geometry.

\section{\label{sec:Conclusion}Conclusions}

Wigner's theorem on the realization of symmetries in quantum mechanics
is a result of fundamental importance in Physics. Although different
proofs of this theorem are available in the literature, to the best
of our knowledge non of them emphasizes the close relation that exists
between Wigner's theorem and the fundamental theorem of projective
geometry. In this paper, we have adopted a strictly geometric point
of view in order to prove Wigner's theorem. Using the tools of projective
geometry, suitably adapted to the case of infinite dimensions, Wigner's
theorem becomes a corollary of the fundamental theorem of projective
geometry. In particular, we have proved that any symmetry transformation
is a projective collineation and hence (by the fundamental theorem
of projective geometry) must also be a semi-projectivity, i.e. it
is induced by a semi-linear (linear or anti-linear) map on the Hilbert
space. Using the fact that symmetry transformations preserve, by definition,
transition probabilities, we have also proved that this semi-linear
map is in fact either a unitary or an anti-unitary operator. As stated
in the introduction, there are some proofs that mention the relation
to projective geometry. But the close connection to the fundamental
theorem of projective geometry is obscured by the parallel use of
other concepts. In reference \cite{Varadarajan1968}, projective
geometry appears as a structure naturally contained in a system of
propositions describing a quantum system. In that context, symmetries
are interpreted as automorphisms of the proposition structure, and
a link to Wigner's theorem can be obtained, though not a direct one,
in the sense that many other, rather difficult, concepts are involved.
A similar remark applies to \cite{Emch1963}. In \cite{Lomont1963}, the similarity between
 Wigner's theorem
and Artin's proof \cite{Artin1957} of the fundamental theorem of projective
geometry is mentioned, but not fully exploited. In contrast, in our
analysis we use the tools of projective geometry throughout. This
has enabled us to obtain Wigner's theorem in a clear and concise
way as a consequence of the fundamental theorem of projective geometry.

\section*{acknowledgements} We thank Florian Scheck for the helpful
reading of the manuscript. N.P. would also like to thank Stefan
M\"uller-Stach for interesting discussions. A. F. Reyes-Lega
acknowledges financial support from the Faculty of Sciences of
Universidad de los Andes. K. Keller thanks the DAAD for financially
supporting the joint work with A. F. Reyes-Lega.



\end{document}